\documentclass[conference]{IEEEtran}
\pdfoutput=1 %flag for PDFLaTex processing by arXiv
\usepackage{cite}
\usepackage{amsmath,amssymb,amsfonts}
\usepackage{algorithmic}
\usepackage{graphicx}
\usepackage{textcomp}
\usepackage{xcolor}
\def\BibTeX{{\rm B\kern-.05em{\sc i\kern-.025em b}\kern-.08em
    T\kern-.1667em\lower.7ex\hbox{E}\kern-.125emX}}

\usepackage[utf8]{inputenc} %encoding
\usepackage{bbold}

\usepackage{epstopdf}

\usepackage{units}
\usepackage{bm} % bold 

\usepackage{array}
\usepackage{booktabs} % for much better looking tables
\usepackage{tabularx} % formatted tables
\usepackage{multirow}
\usepackage{multicol}
\usepackage{adjustbox}
\newcolumntype{L}[1]{>{\hsize=#1\hsize\raggedright\arraybackslash}X} % input document customizations
\begin{document}

\title{\huge RTO Frequency Regulation Market Clearing Formulations and the Effect of Opportunity Costs on Electricity Prices}

\author{\IEEEauthorblockN{Adria E. Brooks}
\IEEEauthorblockA{\textit{Department of Electrical and Computer Engineering} \\
\textit{University of Wisconsin--Madison}\\
Madison, USA \\
brooks7@wisc.edu}
\and
\IEEEauthorblockN{Bernard C. Lesieutre}
\IEEEauthorblockA{\textit{Department of Electrical and Computer Engineering} \\
\textit{University of Wisconsin--Madison}\\
Madison, WI\\
lesieutre@wisc.edu}}

\maketitle

% --- % Abstract % --- %
\begin{abstract}

The Regional Transmission Operators each have unique methods to procure frequency regulation reserves used to track real power fluctuations on the grid. 
Some market clearing practices result in high regulation prices with large spread compared to other markets.
We present the frequency regulation market clearing formulations--as derived from operator tariffs--of the ISO New England, PJM Interconnection, and Midcontinent ISO.
We offer test case examples to explain the historic market pricing behavior seen within each system operator and conclude this behavior is due to the inclusion of estimated lost opportunity costs in market clearing. 
\end{abstract}

\begin{IEEEkeywords}
RTO/ISO, power system economics, system dispatch, frequency regulation, ancillary services, opportunity costs
\end{IEEEkeywords}

% --- % Introduction % --- %
\section{Introduction} \label{sec:introduction}
%% Introduction %%

Frequency regulation reserves are used to correct the short-term frequency imbalances which occur during the normal operation of the grid. Frequency regulation is the injection or withdrawal of real power by facilities capable of responding to a system operator's automatic generation control (AGC) signal. 

The seven Regional Transmission Operators (RTO) / Independent System Operators (ISO) in the United States each maintain a wholesale market to compensate resources for providing frequency regulating reserves. There are two regulation market clearing prices (RMCP) calculated: capacity (RMCCP) and performance (RMPCP). Regulation capacity ensures a resource is available to provide this service while regulation performance quantifies a resource's movement in response to AGC signals.

\begin{figure}[htbp]
	\centering
	\includegraphics[width=1.0\linewidth]{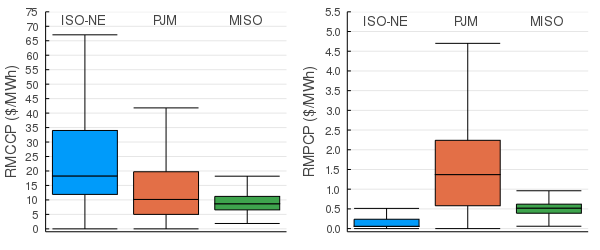} %{clearingPrices.png}
	\caption{2018 hourly regulation market capacity (RMCCP) and performance (RMPCP) clearing prices and ISO-NE, PJM, and MISO by season. Boxes include first, second and third quartiles, whiskers include 1.5 IQR.} %Figure duplicated from \cite{Brooks19_regReview}.}
	\label{fig:clearingPrices}
\end{figure}

We presented a review of the ISO New England (ISO-NE), the PJM Interconnection (PJM), and the Midcontinent ISO (MISO) in \cite{Brooks19_regReview}. That work included the statistics of the 2018 hourly RMCPs for the three operators, shown in Fig. \ref{fig:clearingPrices}. We hypothesized that the large spread seen in some market prices is related to the inclusion of estimated lost opportunity costs in the market clearing. This work complements that review by presenting the regulation market clearing formulations of the three RTOs and explores reasons for the pricing behavior.

% --- % Formulations % --- %
\section{Market Clearing Formulations}
% --- % Market Clearing Formulations % --- %

This work presents regulation market clearing formulations for each RTO as a means to compare the three markets. Most RTOs do not publish their clearing formulations. All formulations presented here are derived from the legal descriptions in each RTO's Business Practice Manuals and Tariffs. The system operators each have unique ways of optimizing regulation in regards to energy and contingency reserves, so these products are included in the market formulations. Transmission network equations are ignored in these formulations. More detailed descriptions of the electricity products are in \cite{Brooks19_regReview}.

Though all markets have unique naming conventions and market constraints, we use consistent variables and nomenclature when possible. The unique naming convention of each RTO is identified in the corresponding section. %All variables and parameters used in market clearing formulations Eqs. \ref{eq:isone_reg} -- \ref{eq:miso_rmcp} are listed below. 

\begin{footnotesize} 
\vspace{0.3cm}
\noindent \textbf{Nomenclature}

\vspace{0.1cm}
\noindent \textit{System Parameters (MW)} \\
\begin{tabular}{p{0.18\columnwidth}<{\raggedleft}p{0.74\columnwidth}}
	$\underline{P}$, $\bar{P}$ & minimum and maximum resource limits \\
	$\bar{P}_{d}$ & energy demand requirement \\
	$\bar{R}_{cap}$, $\bar{R}_{per}$ & regulation capacity and performance requirements \\
	$\bar{R}_{10}$, $\bar{R}_{30}$ & 10- and 30-min contingency reserve requirements \\
	$\bar{R}_{syn}$, $\bar{R}_{sup}$ & Synchronous and supplemental reserve requirements \\
\end{tabular}

\vspace{0.2cm}
\noindent \textit{Resource Parameters (\$/MWh)} \\
\begin{tabular}{p{0.18\columnwidth}<{\raggedleft}p{0.74\columnwidth}}
    $c_p$ & offer price for energy \\
    $c_{cap}$, $c_{per}$ & offer price for regulation capacity (\$/MWh) and performance (\$/MW) \\
    $c_{syn}$, $c_{non}$, $c_{sup}$ & offer price for synchronous, non-synchronous, and supplemental reserves \\
    $\tilde{c}_{loc}$, $c_{loc}$ & estimated and real-time incremental opportunity cost \\
    $\pi$ & incremental avoided cost  \\
    $V$ & Vickrey payment (\$)  \\
    $R_o$ & maximum regulation capacity offer (MW)  \\
    $\rho$ & ramp rate of resource (MW/min) \\
    $t$ & length of dispatch interval (min) \\
    $\alpha$ & historic mileage ratio or deployment performance (1/hr) \\
\end{tabular}

\vspace{0.5cm}
\noindent \textit{Dispatch Variables (MW)} \\
\begin{tabular}{p{0.18\columnwidth}<{\raggedleft}p{0.74\columnwidth}}
	$P$ & energy dispatch \\
	$R_{cap}$, $R_{per}$ & regulation capacity and performance dispatch \\
	$R^1$ & regulation capacity dispatch as regulation \\
	$R^2$ & regulation capacity dispatch as operating reserve \\
	$R_{syn}$, $R_{non}$, $R_{sup}$ & synchronous, non-synchronous, non-synchronous reserve dispatch \\
\end{tabular}

\vspace{0.3cm}
\noindent \textit{Clearing Prices (\$/MWh)} \\
\begin{tabular}{p{0.18\columnwidth}<{\raggedleft}p{0.74\columnwidth}}
	$\tilde{\gamma}$, $\gamma$ & estimated and real-time locational marginal price \\
	$\mu_{cap}$, $\mu_{per}$, $\mu$ & regulation capacity, performance, and total market clearing price \\
\end{tabular}

\vspace{0.2cm}
\noindent \textit{Sets} \\
\begin{tabular}{p{0.18\columnwidth}<{\raggedleft}p{0.74\columnwidth}}
	$i$ & all resources \\
	$j$ & resources providing regulation ($j \in i$) \\
	$k,l,m$ & resources providing synchronous, non-synchronous, supplemental reserves ($k,l,m \in i$) \\
	$f$ & resources in forward market, or previous step ($f \in i$)
\end{tabular}

\end{footnotesize}

% --- % ISO-NE % --- %
%\vspace{1cm}
\subsection{ISO New England} \label{sec:ISONE}
% - % SECTION: ISO-NE % -- %

%%
\subsubsection{Market Summary}

ISO New England has ancillary services for regulation, 10-minute spinning (synchronous) reserves, 10-minute non-synchronous reserves, and 30-minute operating (supplemental) reserves. ISO-NE treats its two regulation components--\textit{capacity} and \textit{service}--as independent market products, each of which must be dispatched to meet an independent system requirement based on historic performance \cite{ISONE18_mr1}.

In real-time, all three contingency reserve products are co-optimized with energy, but regulating reserves are procured independently one hour ahead of the real-time energy and contingency reserve market. ISO-NE selects regulation products based on economic merit order considering a two-part regulation offer and estimated incremental opportunity costs. Resource-specific incremental opportunity costs are estimated for each eligible resource using estimated energy market conditions before real-time \cite{Henson19_convo}.

The regulation service clearing price is simply the most expensive bid from all accepted regulation offers \cite{ISONE18_mr1}.

%% Note: this paragraph C&P from EJ
The regulation capacity clearing price is set using a Vickrey auction design. According to this design, a resource's regulation Vickrey payment should include it's cost of providing regulation capacity, cost of providing regulation service, incremental opportunity cost and the incremental avoided cost the resource provides to the system \cite{Cramton12_FERCtestimony}. The latter can be calculated explicitly by solving the frequency regulation market both with and without the individual resource and taking the difference in the objective value \cite{ISONE18_mr1}, as shown in Eq. (\ref{eq:avoidedCost}) for resource $j$. The optimal Vickrey payment owed to resource $j$ is shown in Eq. (\ref{eq:VickreyPayment}). The regulation capacity market clearing price is set equal to the residual of total Vickrey payment after subtracting estimated mileage payments \cite{Cramton12_FERCtestimony}.

\vspace{0.1cm}
% \begin{small}
\begin{footnotesize}
\noindent \textbf{Incremental Avoided Cost:}
\begin{equation}
	%\text{Incremental Avoided Cost: }
	\pi_j = f^{\ast} (j \notin i) - f^{*} (j \in i) 
	\label{eq:avoidedCost}	
\end{equation}

\noindent \textbf{Vickrey Payment:}
\begin{equation}
	%\text{Vickrey Payment: }
	V_{j} = (c_{cap,j} + c_{loc,j}) \cdot R_{cap,j} + c_{per,j} \cdot R_{per,j} +\pi_j 
	\label{eq:VickreyPayment}	
\end{equation}
% \end{small}
\end{footnotesize}

%--% Subsection
\subsubsection{Market Clearing Formulation}

The regulation capacity dispatch resulting from hourly regulation market clearing is used as an input into the real-time energy and operating reserve markets for the next hour. The expected regulation service is considered an optimization variable in the regulation market.

The regulation market clearing formulation is presented in Eq. (\ref{eq:isone_reg}). Constraints (\ref{eq:isone_reg}a) - (\ref{eq:isone_reg}c) define the range and ramp constraints on both regulation components. $\beta$ in (\ref{eq:isone_reg}b) is the maximum deployment ratio for regulation service. This could be resource specific, but is--
at most--the product of twice the total $R_{cap}$ range and the number of AGC signals sent within the regulation dispatch interval \cite{chen15_2partRegulation}. Constraints (\ref{eq:isone_reg}d) and (\ref{eq:isone_reg}e) define the system regulation capacity and service requirements. The estimated incremental opportunity cost is defined in (\ref{eq:isone_reg}f) using a forecasted energy price \cite{Henson19_convo}.

The real-time energy and reserve market clearing formulation is presented in Eq. (\ref{eq:isone_energy}). This market takes the regulation capacity dispatch ($R_{cap}$) from Eq. (\ref{eq:isone_reg}) as a fixed input in power generation constraints (\ref{eq:isone_energy}a) and (\ref{eq:isone_energy}b). Constraints (\ref{eq:isone_energy}c) - (\ref{eq:isone_energy}c) define the system energy, 10-min and 30-min contingency reserve needs, respectively.

The regulation market clearing prices are defined in Eq. (\ref{eq:isone_rmcp}).

% \vfill
\vspace{0.5cm}
% \begin{small}
\begin{footnotesize}
\noindent \textbf{Regulation Market:}
\begin{align}
	& \text{min}
	& & \left( c_{cap} + \tilde{c}_{loc} \right) ^T R_{cap} + c_{per}^T R_{per} \label{eq:isone_reg} \\
	& \text{s.t.}
	& & R_{cap,j} \le \frac{1}{2} \left( \bar{P}_{j} - \underline{P}_{j} \right) \tag{a} \\
	& & & R_{per,j} \le \beta \cdot R_{cap,j}  \tag{b} \\
     & & & R_{per,j} \le t \cdot \rho_{j} \cdot R_{cap,j} \tag{c} \\
	& & & \sum_j R_{cap,j} \ge \bar{R}_{cap} \tag{d} \\
	& & & \sum_j R_{per,j} \ge \bar{R}_{per} \tag{e} \\
	& & & R_{cap}, R_{per} \geq 0 \nonumber \\ \nonumber \\
	& & & \tilde{c}_{loc} = \tilde{\gamma} - c_p \tag{f} \\ 
	& & & \text{Dispatched: } R_{cap} \nonumber
\end{align}

%\vfill
%\vspace{0.1cm}
\noindent \textbf{Co-Optimized Energy and Contingency Reserve Market:}
\begin{align}
	& \text{min}
	& & c_p^T P + c_{syn}^T R_{syn} + c_{non}^T R_{non} + c_{sup}^T R_{sup} \label{eq:isone_energy} \\
	& \text{s.t.}
	& & P_i + R_{cap,f} + R_{syn,k} + R_{sup,m} \leq \bar{P}_{i} \tag{a} \\
	& & & P_i - R_{cap,f} \geq \underline{P}_{i} \tag{b} \\
	& & & \sum_i P_i = \bar{P}_{d} \tag{c} \\
	& & & \sum_f R_{non,f} + \sum_l R_{non,l}  + \sum_k R_{syn,k} \geq \bar{R}_{10} \tag{d} \\
	& & & \sum_f R_{sup,f} + \sum_m R_{sup,m} \geq \bar{R}_{30} \tag{e} \\
	& & & P, R_{syn}, R_{non}, R_{sum} \geq 0 \nonumber \\
	\nonumber \\
	& & & \text{Dispatched: } P, R_{syn}, R_{non}, R_{sup} \nonumber
\end{align}

%\vfill
%\vspace{0.1cm}
\noindent \textbf{Market Prices:}
\begin{align} 
	& \mu = \mu_{per} + \mu_{cap} \label{eq:isone_rmcp} \\
	& \mu_{per} = \max \{c_{per,j} \} \; \text{for} \; R_{per,j} \neq 0 \tag{a} \\
	& \mu_{cap} = \frac{\sum_j V_j - \sum_j R_{per,j} \cdot \mu_{per} }{\sum_j R_{cap,j}} \tag{b}
\end{align}
% \end{small}
\end{footnotesize}

% --- % PJM % --- %
%\vspace{1cm}
\subsection{PJM Interconnection} \label{sec:PJM}
% - % SECTION: PJM % - %

%%
\subsubsection{Market Summary}
PJM has ancillary service products for regulating reserves, 10-minute synchronous reserves, 10-minute non-synchronous reserves, and 30-minute supplemental reserves. 
Regulation reserves are procured on the operating day one hour ahead of dispatch. Though not independent regulation products, PJM splits regulation into two components--\textit{capability} and \textit{performance}. 

PJM clears energy and ancillary services using a multi-step process. The two relevant steps for this work are the Ancillary Service Optimizer (ASO) and the Real-Time Security Constrained Economic Dispatch (RT SCED).
The ASO jointly optimizes energy and all reserves in order to commit regulating and non-dispatchable contingency reserves \cite{PJMpres_regulationIntroOffersClearing}. %An estimated incremental regulation opportunity cost -- the difference between the day-ahead LMP and the energy offer at the regulation setpoint -- is calculated for each resource providing regulation \cite{PJMpres_regulationOverview, PJM16_regUpliftLOC}. 
An estimated opportunity cost is considered in the ASO to clear the market and the resulting regulation dispatch commitments are used as inputs in the RT SCED \cite{PJMpres_reserveClearingPricing}. 

%%%
\subsubsection{Market Clearing Formulation}

Eq. (\ref{eq:PJM_ASO}) shows the ASO step. The objective function minimizes the cost of regulation including the capability offers, the performance offers scaled by the mileage ratio, and an estimated incremental lost opportunity cost using the day-ahead energy LMP. Constraints (\ref{eq:PJM_ASO}a) and (\ref{eq:PJM_ASO}b) define the power dispatch limits. Constraint (\ref{eq:PJM_ASO}c) limits regulation capacity dispatch to the resource specified limit. Constraints (\ref{eq:PJM_ASO}d) - (\ref{eq:PJM_ASO}g) define the system requirements for energy demand, regulation capacity, synchronous reserves, and 10-minute contingency reserves, respectively. The incremental opportunity cost based on day-ahead energy estimates is defined in Eq. (\ref{eq:PJM_ASO}h).

Eq. (\ref{eq:PJM_RTSCED}) shows the RT SCED step. Constraints (\ref{eq:PJM_RTSCED}a) and (\ref{eq:PJM_RTSCED}b) define the power dispatch limits. Constraints (\ref{eq:PJM_RTSCED}c) - (\ref{eq:PJM_RTSCED}e) define the system requirements for energy demand, synchronous reserves, and 10-minute contingency reserves, respectively.

% \vfill
\vspace{0.5cm}
% \begin{small}
\begin{footnotesize}
\noindent \textbf{Ancillary Service Optimizer:}
\begin{align}
	& \text{min}
	& & c_p^T P + \left( c_{cap} + \alpha \cdot c_{per} + \tilde{c}_{loc} \right)^T R_{cap} + c_{syn}^T R_{syn} + c_{non}^T R_{non}
\label{eq:PJM_ASO} \\
	& \text{s.t.}
	& & P_i + R_{cap,j} + R_{syn,f} + R_{sup,f} + R_{syn,k} \leq \bar{P}_{i} \tag{a} \\
	& & & P_i - R_{cap,j} \geq \underline{P}_{i} \tag{b} \\	
    & & & R_{cap,j} \leq R_{o,j} \tag{c} \\
	& & & \sum_i P_i = \bar{P}_{d}	\tag{d} \\
	& & & \sum_j R_{cap,j} \geq \bar{R}_{cap}  \tag{e} \\
	& & & \sum_f R_{syn,f} + \sum_k R_{syn,k} \geq \bar{R}_{syn}  \tag{f} \\
	& & & \sum_f \left( R_{syn,f} + R_{non,f} \right) + \sum_k R_{syn,k} + \sum_l R_{non,l} \geq \bar{R}_{10} \tag{g} \\	
	& & & P, R_{cap}, R_{syn}, R_{non} \geq 0  \nonumber \\
	\nonumber \\
	& & & \tilde{c}_{loc} = ( \tilde{\gamma} - c_p ) \tag{h}	 \\
	& & & \text{Dispatched: } R_{cap} \nonumber
\end{align}

%\vfill
% \vspace{0.2cm}
\noindent \textbf{Real-Time Security Constrained Economic Dispatch:}
\begin{align}
	& {\text{min}}
	& & c_p^T P + c_{syn}^T R_{syn} + c_{non}^T R_{non}
\label{eq:PJM_RTSCED} \\
	& \text{s.t.}
	& & P_i + R_{cap,f} + R_{syn,f} + R_{sup,f} + R_{syn,k} \leq \bar{P}_{i} \tag{a} \\
	& & & P_i - R_{cap,f} \geq \underline{P}_{i} \tag{b} \\
	& & & \sum_i P_i = \bar{P}_{d}	\tag{c} \\
	& & & \sum_f R_{syn,f} + \sum_k R_{syn,k} \geq \bar{R}_{syn}  \tag{d} \\	
	& & & \sum_f \left( R_{syn,f} + R_{non,f} \right) + \sum_k R_{syn,k} + \sum_l R_{non,l} \geq \bar{R}_{10}  \tag{e} \\	
	& & & P, R_{syn}, R_{non} \geq 0 \nonumber \\
	\nonumber \\
	& & & \text{Dispatched: } P, R_{syn}, R_{non} \nonumber
\end{align}

%\vfill
% \vspace{0.2cm}
\noindent \textbf{Market Prices:}
\begin{align}
	& \mu = \max \{ c_{cap} + \alpha \cdot c_{per} + c_{loc} \} \; \text{for} \, R_{cap,j} \neq 0 \label{eq:pjm_rmcp} \\
	& \mu_{per} = \max \{ c_{per} \} \; \text{for} \, R_{cap,j} \neq 0 \tag{a} \\
	& \mu_{cap} = \mu - \mu_{per} \tag{b} \\
	& c_{loc} = \gamma - c_p \tag{c}
\end{align}
% \end{small}
\end{footnotesize}

Resource-specific incremental lost opportunity cost prices are re-calculated using the real-time LMP for use in pricing \cite{PJMpres_regulationOverview}. The marginal resource sets the RMCP, shown in Eq. (\ref{eq:pjm_rmcp}), and the RMPCP is simply the highest adjusted performance offer of the committed resources, shown in Eq. (\ref{eq:pjm_rmcp}a). Different resources can set the RMCP and RMPCP \cite{PJMpres_regulationOverview}. The RMCCP is the difference between the RMCP and the RMPCP, shown in Eq. (\ref{eq:pjm_rmcp}b). The RMCCP will include both regulation capability offers and incremental opportunity costs.

% --- % MISO % --- %
%\vspace{1cm}
\subsection{Midcontinent ISO} \label{sec:MISO}
% - % SECTION: MISO % - %

%%
\subsubsection{Market Summary}
The Midcontinent ISO energy and reserve products of interest here are energy, regulating reserves, spinning reserves, and supplemental reserves. 
MISO's regulation reserves components are \textit{capacity} and \textit{mileage}. 
Regulation capacity is dispatched to meet pre-determined reserve zone requirements \cite{chen14_zonalRequirements}. 
Regulation mileage is estimated for each generator selected to provide capacity using historic performance information.

MISO clears all energy and reserve products simultaneously both day-ahead (DA) and in real-time (RT) using a co-optimized security-constrained economic dispatch (SCED) formulation. 
The clearing price for each product is the marginal cost of providing that product to the grid, following marginal pricing theory \cite{Galiana05_coupledReserves}. The regulation market clearing price, therefore, guarantees recovery of operating reserve cost offers and the opportunity cost of energy re-dispatch for cleared operating reserves. 
MISO does calculate a separate regulation mileage price, but not a separate regulation capacity price \cite{MISO_BPM002}.

\subsubsection{Market Clearing Formulation}

A simplified real-time clearing formulation for MISO is shown in Eq. (\ref{eq:MISO_clearing}). %This formulation ignores zonal requirements and the penalty demand curves MISO applies to price reserve products in the event that reserve requirements are not met \cite{MISO18_schedule29A, MISO18_schedule28}. 

Because regulation reserves can substitute for contingency reserves, MISO splits regulating reserves into two categories: those dispatched to meet regulation requirements ($R^1$) and those dispatched to meet contingency reserves ($R^2$) \cite{chen15_2partRegulation}. 
Separating regulation in this ways allows for the appropriate resource offers to be applied to each. 
A combined regulation offer is used to select $R^1$ regulating reserves, where the regulation mileage offer is scale by the market-wide, historic deployment performance ratio ($\alpha$) as shown in (\ref{eq:MISO_clearing}h).

Constraints (\ref{eq:MISO_clearing}a) and (\ref{eq:MISO_clearing}b) define the generator limits when providing regulation. Constraint (\ref{eq:MISO_clearing}c) limits the regulating reserves to less than the resource specified offer. 
Constraints (\ref{eq:MISO_clearing}d) - (\ref{eq:MISO_clearing}g) define the system energy demand, regulating reserve requirement, spinning reserve requirement, and total operating reserve requirement, respectively. 

The regulation market clearing price is defined as the marginal cost of providing that reserve product to the system \cite{chen15_2partRegulation}. 
As shown in Eq. (\ref{eq:miso_rmcp}), the RMCP is the sum of the shadow prices ($\eta$) related to meeting regulation, spinning and operating reserve constraints from Eq. (\ref{eq:MISO_clearing}). 
The mileage market price is the highest accepted offer, as shown in (\ref{eq:miso_rmcp}a).

% \vfill
\vspace{0.5cm}
% \begin{small}
\begin{footnotesize}
\noindent \textbf{Security Constrained Economic Dispatch:}
\begin{align}
	& \text{min}
	& & c_p^T P + c_{reg}^T R^1 + c_{cap}^T R^2 + c_{syn}^T R_{syn} + c_{sup}^T R_{sup} \label{eq:MISO_clearing} \\
	& \text{s.t.}
	& & P_i + R_j^1 + R_j^2 + R_{syn,k} + R_{sup,m} \leq \bar{P}_{i} \tag{a} \\
	& & & P_i - R_j^1 \geq \underline{P}_{i} \tag{b} \\
	& & & R_j^1 + R_j^2 \leq R_o \tag{c} \\
	& & & \sum_i P_i = \bar{P}_{d} \tag{d} \\
	& & & \sum_j R_j^1 \geq \bar{R}_{cap} \tag{e} \\
	& & & \sum_j \left( R_j^1 + R_j^2 \right) +\sum_k R_{syn,k} \geq \bar{R}_{cap} + \bar{R}_{syn} \tag{f} \\
	& & & \sum_j \left( R_j^1 + R_j^2 \right) +\sum_k R_{syn,k} + \sum_m R_{sup,m} \geq \nonumber \\  %break up
	& & & \qquad\qquad\qquad\qquad\qquad\qquad \bar{R}_{cap} + \bar{R}_{syn} + \bar{R}_{sup} \tag{g} \\ %this is continuation of above
	& & & P,R^1,R^2,R_{syn},R_{sup} \geq 0   \nonumber \\
	\nonumber \\
	& & & c_{reg} = c_{cap} + \alpha \cdot c_{per} \tag{h} \\
	& & & \text{Dispatched: } P,R^1,R^2,R_{syn},R_{sup} \nonumber 
\end{align}

% \vspace{0.2cm}
\noindent \textbf{Market Prices:}
\begin{align}
	& \mu = \eta_e + \eta_f + \eta_g \label{eq:miso_rmcp}\\
	& \mu_{per} = max \{ c_{per,j} \} \; \text{when} \, R_j^{1} \neq 0 \tag{a}
\end{align}
% \end{small}
\end{footnotesize}

% --- % Clearing Comparison % --- %
%\vspace{1cm}
\section{Market Clearing Test Case} \label{sec:testcase}
%%% Test Case Comparison

In \cite{Brooks19_regReview} we hypothesized that the relatively large spread in some regulation clearing prices seen in Fig. \ref{fig:clearingPrices} were due to the inclusion of an estimated incremental lost opportunity cost. If this value was determined using a poor forecast of the real-time system demand needs, a suboptimal dispatch could occur. 

We test this hypothesis using the regulation market clearing formulations for each RTO with different forecasted and real-time power demand conditions. We present first a small 5-generator test case to show the effects of a poor demand forecast on both dispatch and clearing prices. We then present the clearing price statistics for simulated random demand conditions of a 50-generator economic dispatch example.

% --- % 5-Gen Example Market Clearing % --- %
\subsection{Five Generator Example}
\label{sec:exDispatch}

\begin{table}[htbp]
\caption{Characteristics of five generators used in the market clearing examples.}
\begin{tabularx}{0.49\textwidth}{@{} L{2.0} L{0.8} L{0.8} L{0.8} L{0.8} L{0.8} @{}}
\toprule
	Generator: & A & B & C & D & E     \\ \midrule
	$\rho$ (MW/min) & 5 & 6 & 1 & 10 & 5  \\ 
	$c_p$ (\$/MWh) & 10.00 & 20.00 & 15.00 & 18.00 & 12.00 \\
	$c_{cap}$ (\$/MWh) & 2.50 & 15.00 & 6.50 & 12.00 & 7.00 \\
	$c_{per}$ (\$/MWh) & 1.00 & 1.50 & 0.50 & 2.00 & 2.00 \\
\bottomrule
\end{tabularx}
\label{tab:exDispatch}
\end{table}

Two example economic dispatches using different demand forecasts are presented here. The presented clearing formulations are applied to a five generator example. We clear the same market test case when the forecasted conditions match the real-time conditions and when they differ.

Five generators with power dispatch capabilities between 0 and 100 MW are available to serve system demand and required regulation. 
The ramp rates and cost offers of the five example generators are shown in Table \ref{tab:exDispatch}. 
All generators are available to provide 50 MW of regulation capacity and respond to 4-sec AGC signals.
The system requires 420 MW of power demand and 25 MW of required regulation capacity. 
The hourly required system mileage is set to 80 MW, either explicitly in the case of ISO-NE or by setting the historic mileage ratio ($\alpha$) to 3.2 in the case of PJM and MISO.

% - % Dispatch Example 1 % - %
\subsubsection{Matched $\bar{P_d}$ Conditions}
For the first dispatch example, the same market conditions were used in the forecasted energy clearing as in the real-time. This ensures that the estimated incremental opportunity costs ($\tilde{c}_{loc}$) used in market clearing by ISO-NE and PJM match the real-time incremental opportunity costs used in market pricing. The resulting generator dispatch instructions and clearing prices are shown in Table \ref{tab:dispatchResults}. The assigned mileage ($R_{per}$) in ISO-NE represents dispatch instructions. The assigned mileage ($R_{per}^{*}$) in PJM and MISO represents an assumed mileage if the resources dispatched for regulation capacity move at the assigned mileage ratio of 3.2.

ISO-NE and MISO have very similar energy and regulation capacity dispatches. ISO-NE differs from MISO only to meet their independent regulation service requirement, choosing the generator with the lowest performance offer (Generator C) to provide the majority of mileage. ISO-NE clears the lowest prices of all three RTOs in this example.

PJM dispatches the most expensive generator (Generator B) to provide the majority of regulation, maximizing the amount of energy provided by less expensive generators. PJM's clearing method minimizes generator lost opportunity costs without consideration to avoided costs. PJM makes a suboptimal dispatch choice in this example, increasing both regulation clearing prices. 

\begin{table}[htbp]
\caption{Generator dispatch and market clearing prices given forecasted energy conditions that match real-time conditions.}
\begin{tabularx}{0.49\textwidth}{@{} L{2} L{0.76} L{0.81} L{0.81} L{0.81} L{0.81} @{}}
\toprule
	\multicolumn{6}{c}{\textbf{ISO New England}} \\
	\multicolumn{2}{c}{$\gamma$ = 20.00 \$/MWh} & 
	\multicolumn{2}{c}{$\mu_{cap}$ = 13.73 \$/MWh} &
	\multicolumn{2}{c}{$\mu_{per}$ = 1.00 \$/MWh} \\ \midrule
		 & A & B & C & D & E     \\ \cline{2-6}
	$P$ (MW) & 80.00 & 45.00 & 95.00 & 100.00 & 100.00  \\ 
	$R_{cap}$ (MW) & 20.00 & 0 & 5.00 & 0 & 0 \\
	$R_{per}$ (MW) & 1.67 & 0 & 5.00 & 0 & 0 \\ 	 
\toprule
	\multicolumn{6}{c}{\textbf{PJM Interconnect}} \\
	\multicolumn{2}{c}{$\gamma$ = 20.00 \$/MWh} & 
	\multicolumn{2}{c}{$\mu_{cap}$ = 18.30 \$/MWh} &
	\multicolumn{2}{c}{$\mu_{per}$ = 1.50 \$/MWh} \\ \midrule
		 & A & B & C & D & E     \\ \cline{2-6}
	$P$ (MW) & 100.00 & 25.00 & 95.00 & 100.00 & 100.00  \\ 
	$R_{cap}$ (MW) & 0 & 20.00 & 5.00 & 0 & 0 \\
	$R_{per}^{*}$ (MW) & 0 & 5.33 & 1.33 & 0 & 0 \\ 
\toprule
	\multicolumn{6}{c}{\textbf{Midcontinent ISO}} \\
	\multicolumn{2}{c}{$\gamma$ = 20.00 \$/MWh} & 
	\multicolumn{2}{c}{$\mu$ = 15.70 \$/MWh} &
	\multicolumn{2}{c}{$\mu_{per}$ = 1.00 \$/MWh} \\ \midrule
		 & A & B & C & D & E     \\ \cline{2-6}
	$P$ (MW) & 80.00 & 45.00 & 95.00 & 100.00 & 100.00  \\ 
	$R_{cap}$ (MW) & 20.00 & 0 & 5.00 & 0 & 0 \\
	$R_{per}^{*}$ (MW) & 5.33 & 0 & 1.33 & 0 & 0 \\ 
\bottomrule
\end{tabularx}
\label{tab:dispatchResults}
\end{table}

% - % Dispatch Example 2 % - %
\subsubsection{Mismatched $\bar{P_d}$ Conditions}
In this second dispatch example, the forecasted system power demand is only 40\% of the real-time power demand condition. This demand forces different binding constraints and subsequently lowers the forecasted energy clearing price and estimated incremental opportunity costs ($\tilde{c}_{loc}$) used in market clearing by ISO-NE and PJM. All other system conditions are unchanged.
The resulting generator dispatch instructions and clearing prices are shown in Table \ref{tab:dispatchResultsDA}. 

Both ISO-NE and PJM make regulation dispatch decisions based on the poorly estimated incremental opportunity costs and the resulting energy and regulation dispatches differ from Table \ref{tab:dispatchResults} when the day-ahead energy conditions matched real-time conditions for ISO-NE and PJM. 
Most notably, the ISO-NE regulation capacity clearing price and both PJM regulation clearing prices all decreased significantly. MISO's dispatch instructions and clearing prices remained unchanged between the two dispatch examples.

 \begin{table}[htbp]
 \caption{Generator dispatch and market clearing prices given a mismatch in real-time and forecasted conditions. The power demand forecast was 40\% of real-time demand, resulting in low incremental opportunity cost estimates ($\tilde{c}_{loc}$).}
 \begin{tabularx}{0.49\textwidth}{@{} L{2} L{0.76} L{0.81} L{0.81} L{0.81} L{0.81} @{}}
 \toprule
 	\multicolumn{6}{c}{\textbf{ISO New England}} \\
 	\multicolumn{2}{c}{$\gamma$ = 20.00 \$/MWh} & 
 	\multicolumn{2}{c}{$\mu_{cap}$ = 6.93 \$/MWh} &
 	\multicolumn{2}{c}{$\mu_{per}$ = 1.00 \$/MWh} \\ \midrule
 		 & A & B & C & D & E     \\ \cline{2-6}
 	$P$ (MW) & 75.03 & 45.00 & 99.97 & 100.00 & 100.00  \\ 
 	$R_{cap}$ (MW) & 24.97 & 0 & 0.03 & 0 & 0 \\
 	$R_{per}$ (MW) & 1.67 & 0 & 5.00 & 0 & 0 \\ 	 
 \toprule
 	\multicolumn{6}{c}{\textbf{PJM Interconnect}} \\
 	\multicolumn{2}{c}{$\gamma$ = 20.00 \$/MWh} & 
 	\multicolumn{2}{c}{$\mu_{cap}$ = 14.70 \$/MWh} &
 	\multicolumn{2}{c}{$\mu_{per}$ = 1.00 \$/MWh} \\ \midrule
 		 & A & B & C & D & E     \\ \cline{2-6}
 	$P$ (MW) & 80.00 & 45.00 & 95.00 & 100.00 & 100.00  \\ 
 	$R_{cap}$ (MW) & 20.00 & 0 & 5.00 & 0 & 0 \\
 	$R_{per}^{*}$ (MW) & 5.33 & 0 & 1.33 & 0 & 0 \\ 
 \toprule
 	\multicolumn{6}{c}{\textbf{Midcontinent ISO}} \\
 	\multicolumn{2}{c}{$\gamma$ = 20.00 \$/MWh} & 
 	\multicolumn{2}{c}{$\mu$ = 15.70 \$/MWh} &
 	\multicolumn{2}{c}{$\mu_{per}$ = 1.00 \$/MWh} \\ \midrule
 		 & A & B & C & D & E     \\ \cline{2-6}
 	$P$ (MW) & 80.00 & 45.00 & 95.00 & 100.00 & 100.00  \\ 
 	$R_{cap}$ (MW) & 20.00 & 0 & 5.00 & 0 & 0 \\
 	$R_{per}^{*}$ (MW) & 5.33 & 0 & 1.33 & 0 & 0 \\ 
 \bottomrule
 \end{tabularx}
 \label{tab:dispatchResultsDA}
 \end{table}

\subsection{Fifty Generator Example}

This same method was applied to a larger system case using 8760 trials to determine the spread in regulation clearing prices that result. The forecasted energy demand was allowed to fluctuate between 50\% and 125\% of the real-time demand for a  50-generator system. All other inputs were unchanged between the 8760 trials. The resulting statistics for these trials are presented in Table \ref{tab:dispTrials}.

The only clearing prices which experience any variance between trials are the RMCCP for ISO-NE and PJM, and the RMPCP for PJM. This corroborates the findings in the 5-generator example and matches the behavior seen in Fig. \ref{fig:clearingPrices}.

\begin{table}[htbp]
\caption{Statistics of market clearing prices for each RTO given a 50-generator system with 8760 trials. The forecasted energy demand was the only input permitted to fluctuate between trials. Energy price statistics were the same for all RTOs.}
\centering
\begin{tabularx}{0.45\textwidth}{@{} L{1} L{1} L{1} L{1} L{1} L{1} @{}}
\toprule
	 &  & minimum & mean & maximum & variance     \\ \midrule
	\multirow{2}{*}{Energy} & $\tilde{\gamma}$ & 30.50 & 40.69 & 60.00 & 56.98  \\ 
	 & $\gamma$ & 41.00 & 41.00 & 41.00 & 0  \\ \midrule
	\multirow{2}{*}{ISONE} & $\mu_{cap}$ & 4.31 & 4.68 & 6.59 & 0.45 \\
	 & $\mu_{per}$ & 1.00 & 1.00 & 1.00 & 0 \\ \midrule
	\multirow{2}{*}{PJM} & $\mu_{cap}$ & 16.75 & 16.85 & 17.50 & 0.04 \\
	 & $\mu_{per}$ & 2.50 & 3.41 & 4.00 & 0.54 \\ \midrule
	\multirow{2}{*}{MISO} & $\mu_{cap}$ & 19.25 & 19.25 & 19.25 & 0 \\
	 & $\mu_{per}$ & 2.50 & 2.50 & 2.50 & 0 \\
\bottomrule
\end{tabularx}
\label{tab:dispTrials}
\end{table}

% --- % Conclusions % --- %
%\vspace{1cm}
\section{Conclusions} \label{sec:conclusions}
% Conclusion %

The market clearing practices of the ISO New England, PJM Interconnect and Midcontinent ISO differ widely. 
Operators strive to reduce overall system costs given the many energy and reserve products, but the method by which frequency regulating reserves are optimized with energy can result in different dispatches given identical system conditions. 
The inclusion of lost opportunity costs in the market clearing formulations were found to have a large impact on regulation market pricing behavior.
Regulation markets that are cleared before the energy markets must rely on an estimated power demand condition. 
If this demand estimate differs from real-time, then a suboptimal dispatch could occur. 

Historic regulation market clearing prices show a large spread in some prices among the RTOs.
This behavior was replicated using modeled test case systems when the forecasted energy demand differed from real-time conditions.
This supports the observation that suboptimal dispatch decisions could be made on poorly estimated lost opportunity costs. 

The test cases included here did not consider the affect of generator bids on regulation market prices.
Variability in generator energy bids could also contribute to the wide spread of regulation clearing prices. 
% Furthermore, prices can fluctuate under high congestion or contingency events, which tend to drive up all market prices. 
Forecasted power demand is only one component of regulation market clearing prices, and only partially explains historic pricing behavior.

% --- % acknowledgment % --- %
\section*{Acknowledgment}
%\begin{small}
The authors would like to thank Mr. Bill Henson of ISO-NE and Dr. Yonghong Chen of MISO for their guidance in this work. PJM did not return request for review on this work.

This material is based upon work supported by the National Science
Foundation Graduate Research Fellowship Program under Grant No.
DGE-1256259. Any opinions, findings, and conclusions or
recommendations expressed in this material are those of the authors
and do not necessarily reflect the views of the National Science
Foundation. Support was also provided by the Graduate School and the Office of the Vice Chancellor for Research and Graduate Education at the University of Wisconsin-Madison with funding from the Wisconsin Alumni Research Foundation.

A. Brooks acknowledges that this work was performed at the University of Wisconsin-Madison, which occupies ancestral Ho-Chunk land—a place known to their nation as Teejop—which was forcibly ceded in an 1832 treaty. She further acknowledges the invaluable labor of the maintenance and clerical staff at this institution. 
%\end{small}

% Bibliography
%\vspace{1cm}
\bibliography{regMarketBib}

\end{document}